# Complexity without chaos: Plasticity within random recurrent networks generates robust timing and motor control.

Rodrigo Laje[1,2] and Dean V. Buonomano[1*]

[1]Departments of Neurobiology and Psychology and Brain Research Institute, University of California, Los Angeles, CA, USA.

[2]Permanent address: Departamento de Ciencia y Tecnología, Universidad Nacional de Quilmes, Bernal, Argentina, and CONICET, Argentina.

*Correspondence to: dbuono@ucla.edu

It is widely accepted that the complex dynamics characteristic of recurrent neural circuits contributes in a fundamental manner to brain function. Progress has been slow in understanding and exploiting the computational power of recurrent dynamics for two main reasons: nonlinear recurrent networks often exhibit chaotic behavior and most known learning rules do not work in robust fashion in recurrent networks. Here we address both these problems by demonstrating how random recurrent networks (RRN) that initially exhibit chaotic dynamics can be tuned through a supervised learning rule to generate locally stable neural patterns of activity that are both complex and robust to noise. The outcome is a novel neural network regime that exhibits both transiently stable and chaotic trajectories. We further show that the recurrent learning rule dramatically increases the ability of RRNs to generate complex spatiotemporal motor patterns, and accounts for recent experimental data showing a decrease in neural variability in response to stimulus onset.

**INTRODUCTION**

An influential theory in neuroscience is that computations are instantiated by the activity of neural networks converging to steady-state patterns (Amit, 1989; Amit and Brunel, 1997; Brody et al., 2003; Durstewitz et al., 2000; Hopfield, 1982; Wang, 2001). These patterns can be described as fixed-point attractors in neural state space—where each dimension corresponds to the activity level of a neuron and a point in this state space describes a given pattern of activity. Such fixed-point attractor dynamics is likely important for some tasks including working memory. However, more recent theoretical and experimental work has emphasized the importance of dynamic patterns of activity in neural computations (Buonomano and Maass, 2009; Durstewitz and Deco, 2008; Perdikis et al., 2011; Rabinovich et al., 2008; Stopfer et al., 2003). Under this framework changing patterns of neural activity are represented as trajectories in neural state space, and computations arise from the voyage through state space, as opposed to the arrival at any one given location.

There are potentially many advantages of relying on the complex dynamics of recurrent networks to perform computations (Buonomano and Maass, 2009; Durstewitz and Deco, 2008; Jaeger and Haas, 2004; Medina et al., 2000; Rabinovich et al., 2008; Sussillo and Abbott, 2009). Indeed, a number of related theoretical frameworks (sometimes grouped under the term *reservoir computing*) are to a large extent based on the notion that randomly connected recurrent networks (RRNs) can generate patterns of activity in high-dimensional space, and these patterns or trajectories can in turn be used for pattern classification or pattern generation through linear readout units (Buonomano and Maass, 2009; Buonomano and Merzenich, 1995; Jaeger, 2001; Jaeger et al., 2007; Maass et al., 2002; Medina and Mauk, 2000). Consistent with this theoretical work experimental studies indicate that sensory and motor computations may be encoded in the trajectories of neural activity in high-dimensional space (Balaguer-Ballester et al., 2011; Harvey et al., 2012; Mazor and Laurent, 2005; Nikolić et al., 2009; Rabinovich et al., 2001). The advantage of computing with neural trajectories, as opposed to steady states, is particularly obvious for sensory and motor tasks that require timing. Since neural trajectories by their very nature encode time information about stimulus onset, order, and duration, timing is implicitly present. Indeed, models of temporal processing have proposed that the brain encodes time in changing patterns of neural activity (Buonomano and Laje, 2010; Buonomano and Mauk, 1994; Itskov et al., 2011; Mauk and Donegan, 1997; Medina et al., 2000). Within these models any stimulus can elicit a unique neural trajectory, which thus can encode not only the stimulus but the amount of time elapsed since its onset—this information can be used to classify stimuli or generate timed motor responses.

A fundamental challenge, however, in understanding and exploiting the nonlinear dynamics of recurrent networks is that theoretical and experimental results reveal that they often exhibit chaos (Banerjee et al., 2008; Brunel, 2000; Izhikevich and Edelman, 2008; London et al., 2010; Skarda and Freeman, 1987; van Vreeswijk and Sompolinsky, 1996). This is particularly true in "high gain" regimes where the networks are capable of producing aperiodic self-sustained patterns of activity. For example, using a firing rate model Sompolinsky and colleagues elegantly demonstrated that nonlinear networks that exhibited complex dynamics where chaotic (Sompolinsky et al., 1988). Thus minute levels of noise can produce vastly different neural trajectories—effectively abolishing the computational power of a network because a given pattern cannot be reliably reproduced.

A second challenge in understanding and controlling the dynamics of recurrent networks is effective incorporation of plasticity within the recurrent connections. Specifically, both abstract learning rules (e.g., backpropagation) and experimentally derived rules (e.g., STDP, synaptic scaling) are robust primarily in the context of feed-forward circuits, and are often ineffective or unstable when incorporated into recurrent networks in high-gain regimes (Buonomano, 2005; Houweling et al., 2005). Indeed, the nonlinearities and positive feedback characteristic of recurrent networks often renders the incorporation of synaptic plasticity unstable and intractable within recurrent networks (Bengio et al., 1994; Doya, 1992; Pearlmutter, 1995).

Building on models put forth by Jaeger (Jaeger and Haas, 2004) and Sussillo and Abbott (Sussillo and Abbott, 2009) we describe an approach that provides one of the first examples of effective and robust incorporation of plasticity into RRNs in a high gain regime. A novel and powerful computational consequence of this learning rule is that previously chaotic trajectories become locally stable—that is the learning rule creates a locally stable transient channel or "dynamic attractor". Specifically, after training the network exhibits coexisting chaotic and stable trajectories. We show that these stable neural trajectories can dramatically improve the ability of RRNs to tell time and generate complex motor patterns in the presence of high levels of noise. Our results also shed light on a longstanding puzzle in neuroscience: theoretical and experimental studies suggest that the brain exhibits both chaotic and stable dynamic regimes (Balaguer-Ballester et al., 2011; Banerjee et al., 2008; Brunel, 2000; Churchland et al., 2012; Izhikevich and Edelman, 2008; London et al., 2010; Mazor and Laurent, 2005; Rabinovich et al., 2001; Skarda and Freeman, 1987; van Vreeswijk and Sompolinsky, 1996). Our description of a system that has coexisting complex stable transients and chaotic trajectories may reconcile the existence of these two regimes within the brain, and account for recent experimental results demonstrating a decrease in the cross-trial neural variability observed in response to stimulus onset.

## RESULTS

### "Innate" Training

The nonlinear recurrent networks that have been best studied at the theoretical level are RRNs composed of firing rate units with sigmoid activation functions (see Methods). In these networks the connectivity is represented by a weight matrix $\mathbf{W^{Rec}}$ drawn from a normal distribution with a mean of zero and a standard deviation scaled by a parameter $g$ (sometimes referred to as the spectral radius of the connectivity matrix). Sompolinsky and colleagues (Sompolinsky et al., 1988) demonstrated that, for large networks, values of $g>1$ generate increasingly complex and chaotic patterns of activity. **Figure 1A** provides such an example ($g=1.8$, number of units $N=800$). The network is spontaneously active (i.e., it has self-sustaining activity), and an external input at $t=0$ ms (50 ms duration) temporarily kicks the network into a delimited volume of state space, which can be defined as the starting point of a neural trajectory. Across trials, different initial conditions (or the presence of continuous noise, see below) result in a divergence of the trajectories as illustrated in **Figure 1B** (Pre-training) by the firing rates of 3 sample units. This divergence can render the network useless from a computational perspective because the patterns cannot be reproduced across trials. One approach to overcome this problem has been to use carefully tuned feedback to control the dynamics of the network (Jaeger and Haas, 2004; Sussillo and Abbott, 2009). An alternate approach would be to alter the weights of the RRN proper in order to decrease the sensitivity to noise; this approach, however, has been limited by the challenges inherent in changing the weights in recurrent networks. Specifically, since all weights are "being used" throughout the trajectory, plasticity tends to dramatically alter network dynamics, produce bifurcations, and not converge (Doya, 1992; Pearlmutter, 1995).

**Figure 1. Complexity without chaos. A:** A random recurrent network (left panel) in the chaotic regime is stimulated by a brief input pulse (small black rectangle at t=0 in right panel) to produce a complex pattern of activity in the absence of noise. Color-coded raster plot of the activity of 100 out of 800 recurrent units (right panel). Color-coded activity ranges from -1 (blue) to 1 (red). **B:** Time series of three sample recurrent units. In the pre-training panel the blue traces comprised the innate trajectory subsequently used for training. The divergence of the blue and red lines demonstrates that two different initial conditions lead to diverging trajectories before training (left panel), even in the absence of ongoing noise. After training, however, the time series are reproducible during the trained window (2 s; shaded area). That is, despite different initial conditions the blue and red lines trace very similar paths, while still diverging outside of the trained window. **C:** PCA decomposition of the simulations shown in **B**. The input pulse brings different initial conditions into a delimited volume in phase space; after the input is off, the trajectories diverge if the network was not trained (left panel), or trace a very similar path if trained (right panel).

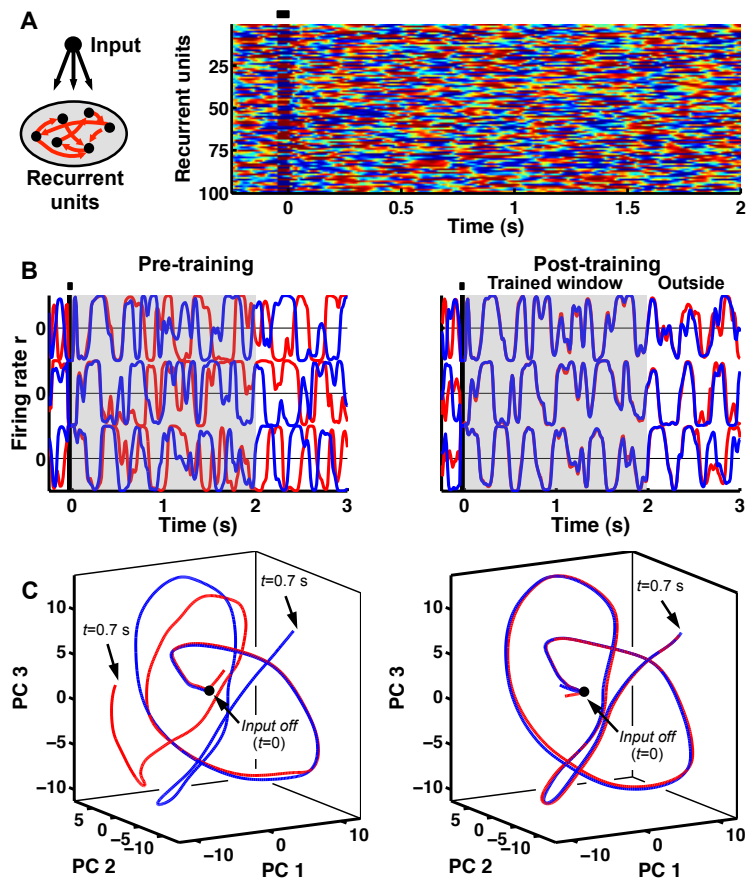

When supervised learning rules are used to train feed-forward or recurrent networks, the traditional approach is to adjust the weights to minimize the error between the actual output units and some desired target. For the reasons mentioned above this approach has proven to be largely intractable when training recurrent networks to learn nonperiodic patterns (Doya, 1992; Pearlmutter, 1995). In the current framework the particular trajectory that the recurrent network uses for a computation is largely irrelevant—what matters is that it is complex and that these patterns can be used by downstream units (Buonomano and Maass, 2009; Jaeger et al., 2007). This means that, independent of the ultimate desired output, there is really no specific desired target activity pattern within the recurrent network.

Thus we reasoned that noise sensitivity could be reduced by training the units in the network to reproduce their "innate" pattern of activity, rather than some trajectory determined by the "desired" output. We define an "innate" trajectory as one triggered by a given input in the absence of noise (using an arbitrary initial condition, and before any training). In other words the approach is to tune the recurrent units to do what they can already do. Towards this end we used the Recursive Least Squares (RLS) learning rule (which we do not consider it to be biologically plausible, see Discussion) and a strategy to rapidly minimize the errors during a trial (Haykin, 2002; Sussillo and Abbott, 2009). By training the RRN to reproduce its innate trajectory over a 2 s period it was possible to create a locally stable transient channel (**Fig. 1B**, Post-training), largely preserving the shape of the original trajectory while turning it into an "attracting" one within the 2-second window. Outside the training window, however, the trajectory rapidly diverges. The first three principal components of the network dynamics (**Fig. 1C**) illustrate that across two trials the trajectory diverges rapidly before training.

**Noise Analysis, Suppression of Chaos, and Stimulus Specificity**
We next examined two critical questions relating to the stability and dynamics of the trained recurrent networks. First, we performed a noise analysis in order to determine if the network could reliably reproduce the trained trajectory in the presence of high levels of noise. To examine this question different levels of noise were continuously injected into all 800 units of the recurrent network. Second, we examined whether training specifically altered the noise sensitivity of the trajectory elicited by the trained input, or if training produced global changes in all network dynamics. This question can be seen as addressing whether learning (creating locally stable trajectories) was stimulus specific. Towards this end each of 10 different networks ($N$=800, $g$=1.8) were stimulated with two different 50 ms long inputs. The neural trajectory produced by Input 1 (In1) served as the "innate" training target (duration of 2 s) for recurrent plasticity, while the trajectory triggered by the second input (In2) served as a "control" to determine the effect of training on non-trained trajectories (**Fig. 2**). Performance was quantified by examining the correlation between the trajectories elicited in the presence of noise in relation to the trajectory in the absence of noise (see Methods).

Over noise amplitudes up to 0.1 performance of the trained networks in response to In1 was essentially perfect. That is, after training and in the presence of noise In1 produced very similar trajectories across trials. Interestingly the reproducibility was fairly specific to the trained pathway. That is, the noise abolished the ability of In2 to reproduce the same trajectory across trials. These results clearly show that training has a stimulus-specific effect on the network dynamics. In this simulations the recurrent training tool place over a relatively small number of trials; sensitivity to noise can be even further decreased by further training (e.g., **Fig. 6**).

**Figure 2. Robustness against noise. A:** Activity of three sample units in the recurrent network at three different levels of noise. Blue: "template" trajectory (no noise); Red: test trajectory" (with noise). The standard deviation of the noise current $I^{noise}$ was 0.001, 0.1, and 1.0 (top to bottom panels; noise amplitude as a fraction of total absolute incoming synaptic weight to each unit averaged across units is 0.007%, 0.7%, and 7%, respectively). Even with a noise amplitude as high as 0.1, the trained network reproduced the target pattern with great accuracy (2 s; shaded area). **B:** Parametric study of robustness against noise (ten different networks). Performance is measured as the averaged Pearson correlation coefficient between model and test trajectories for each condition (after Fisher transformation), mean ± SEM across networks.

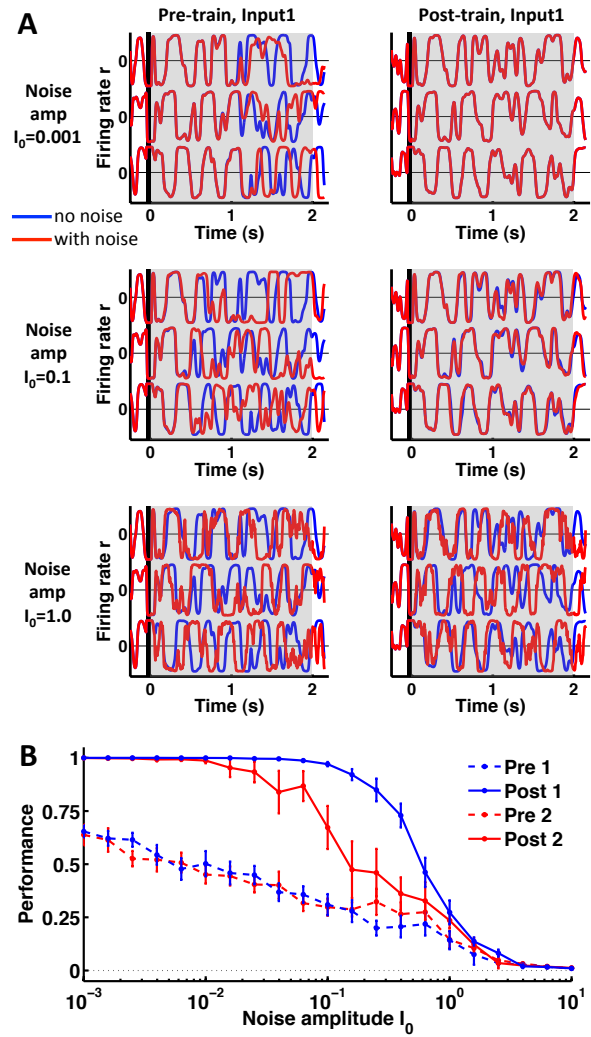

To better characterize the results of the noise analysis and the effects of training we quantified the divergence of trajectories by the largest Lyapunov exponent ($\lambda$), which provides a measure of the rate of separation of two nearby points. For each of the ten networks, $\lambda$ was numerically estimated for the trajectories elicited by In1 and In2, both before and after training (**Fig. 3**). Before training both trajectories exhibited positive exponents, indicative of exponential divergence and thus chaotic dynamics. Overall, training was successful and all networks were able to reproduce the innate trajectory in response to the In1 pulse with great accuracy—despite random initial conditions, the correlation between the test trajectory and innate trajectory was very high ($R$=0.96 ± 0.06, $p$=10$^{-10}$; for In2: $R$=-0.17 ± 0.16, p=0.32; Pearson correlation coefficient over the 2 sec training window averaged after a Fisher transform across units and networks). After training the RRNs, the mean $\lambda$ across networks for In1 was not significantly different from zero, indicative of local stability. The mean $\lambda$ for In2 also decreased, but remained above zero. The dynamics in response to both inputs outside the training window (between $t$=8 s and $t$=10 s) exhibited chaotic dynamics (8/10 networks) or entered stable limit cycles (2/10). Which of these regimes occurred was in part dependent on the initial structure of the network and the extent of the training: lower initial values of $\lambda$ and/or more training loops were more likely to lead to a limit cycle instead of a chaotic attractor (not shown). Importantly, a 2x3 two-way ANOVA with repeated measures (factors "Input" and "Training") showed a significant interaction effect ($p=2\times10^{-5}$), meaning that $\lambda$ post-training was differentially affected by Input1. These results demonstrate that many networks retained a chaotic behavior after innate training, except for the original "innate" trajectory that was transformed into a locally "attracting" trajectory—best described as a "stable transient channel" to the chaotic attractor.

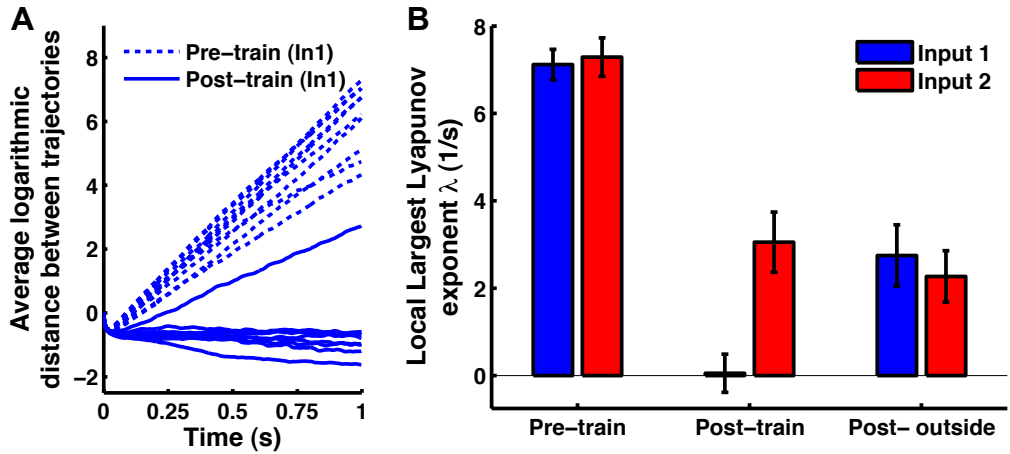

**Figure 3. Suppression of chaos in the trained neural trajectory only. A:** Average logarithmic distance between original and perturbed trajectories for each of ten networks, for the trajectories triggered by Input1 (the trained input). A straight portion with a positive slope is indicative of chaotic dynamics (i.e., the distance between trajectories increases exponentially with time), and the value of the slope is the estimate for the Largest Lyapunov Exponent ($\lambda$). After training, the original and perturbed trajectories are not diverging anymore (except for one network). **B:** The pre-training trajectories triggered by both inputs displayed positive $\lambda$, indicative of chaotic dynamics (Input1: $\lambda=7.12 \pm 0.35$, mean $\pm$ SEM across the ten networks, values significantly different from zero t-test $p=10^{-8}$; Input2: $\lambda=7.29 \pm 0.45$, $p=4\times10^{-8}$; all reported $\lambda$s have units of 1/s). After training, the trajectory triggered by Input1 was locally stable, as indicated by a non-positive mean $\lambda$ ($\lambda=0.05 \pm 0.45$, $p=0.90$); Input2, however, still produced diverging trajectories as evidence by $\lambda$ significantly above zero ($\lambda=3.05 \pm 0.70$, $p=0.0016$). After training the trajectories outside the trained window had a positive mean $\lambda$ in response to both inputs (Input1: $\lambda=2.75 \pm 0.70$, $p=0.0035$; Input2: $\lambda=2.27 \pm 0.60$, $p=0.0039$), with some networks displaying chaotic activity (8/10) and some entering limit cycles (2/10). The interaction effect is significant ($p=2\times10^{-5}$, a 2x3 two-way ANOVA with repeated measures, factors "Input" and "Training"). In addition to this stimulus-specific effect of training, there was a global nonspecific effect of decreased divergence of trajectories after training, represented by a lower though still positive $\lambda$ for Post-train Input2 and Post-outside Input1 and Input2.

## Computational Power of Innate Training

To examine the computational power of the innate training we first quantified the "memory" capacity of the network by determining the maximal delay after the input the network could produce. Towards this end we added an output unit that readout the state of the RRN (**Fig. 4A**). As is typical within the "reservoir" computing framework, the weights of the recurrent units to the output units were adjusted in a supervised fashion. The target output function was flat with a simple pulsed response at different delays after the 50 ms input. **Fig. 4A** shows that a network of size $N=800$ ($g=1.5$) reliably learned a 5000 ms delay, but not a 6000 ms delay, reflecting the finite "memory" of such networks (Ganguli et al., 2008; Jaeger, 2001). To quantify this memory we parametrically varied the delay and compared the performance of the innate training approach to two additional architectures (**Fig. 4B**), using the same set of ten initial networks for all architectures. Together the three architectures were: 1) the current approach ("innate training") where recurrent plasticity within the RRN was directed at the innate trajectory; 2) an echo-state network approach ("echo-state") where the output feeds back onto the RRN, and where only the connections from the recurrent to output units were plastic (Jaeger and Haas, 2004; Sussillo and Abbott, 2009); 3) an RRN with recurrent plasticity ("fair recurrent plasticity", which provided a control for the amount of plastic connections involved in the training); in this architecture the weights of the recurrent units are adjusted according to the error in the output unit (Sussillo and Abbott, 2009, 2012). Both training and testing in this task occurred with random initial conditions and in the presence of noise (noise standard deviation $I_0=0.001$). As shown in **Fig. 4B,** the innate training of the recurrent connections dramatically improved the maximal time delay of the network (defined as the time delay at which performance decays to 0.5), producing on average a 5-fold improvement. Note that the relatively poor performance of the feedback-based architectures is in part because the target is aperiodic—feedback approaches have improved performance when using periodic targets.

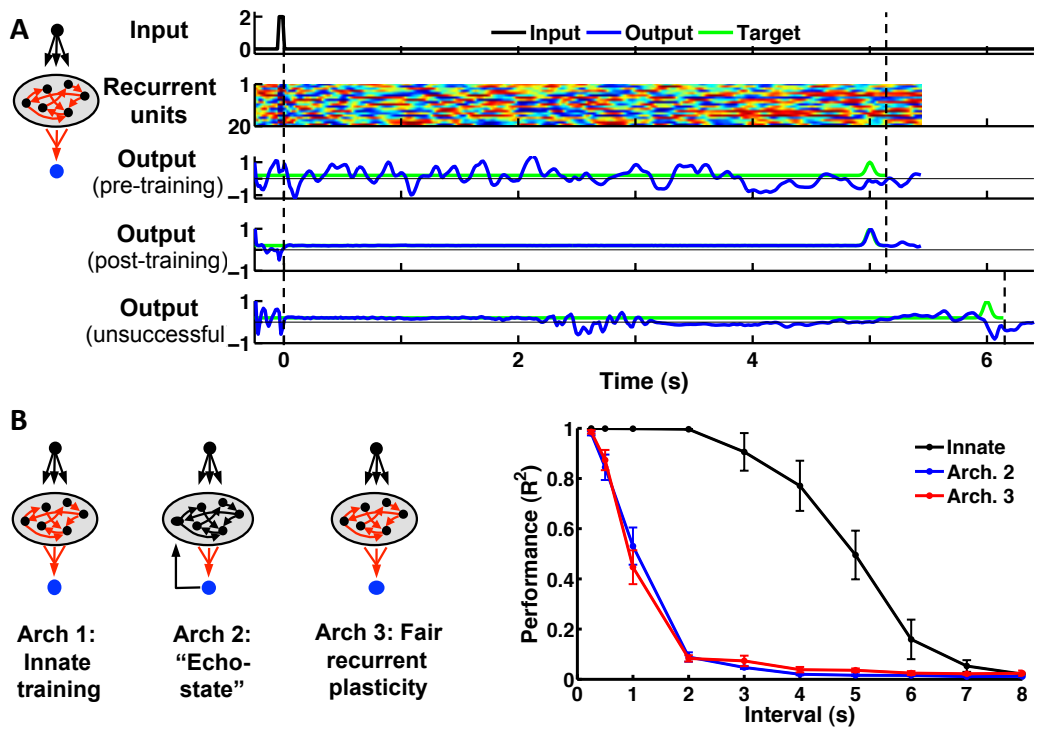

**Figure 4. Enhanced "memory" capacity for timing. A:** An input pulse (black trace) triggers a chaotic innate neural trajectory, displayed as a color-coded raster plot (only 20 out of 800 units shown). A linear readout unit receives input from all the recurrent units (blue trace), showing irregular pre-training activity. After the RRN is trained to the innate trajectory (training window defined by dashed lines), the readout unit is trained to reproduce a flat target with a pulse at a given interval (green trace; 5-s duration in this example). An unsuccessful run from a 6-s interval training is also included as an example. **B:** Performance across different architectures. Ten RRNs were trained in each of the three displayed architectures, parametrically varying the delay. The performance (goodness of reproduction) is quantified by the Pearson correlation coefficient $R^2$ between target and actual output (green and blue traces in **A**); mean ± SEM across networks.

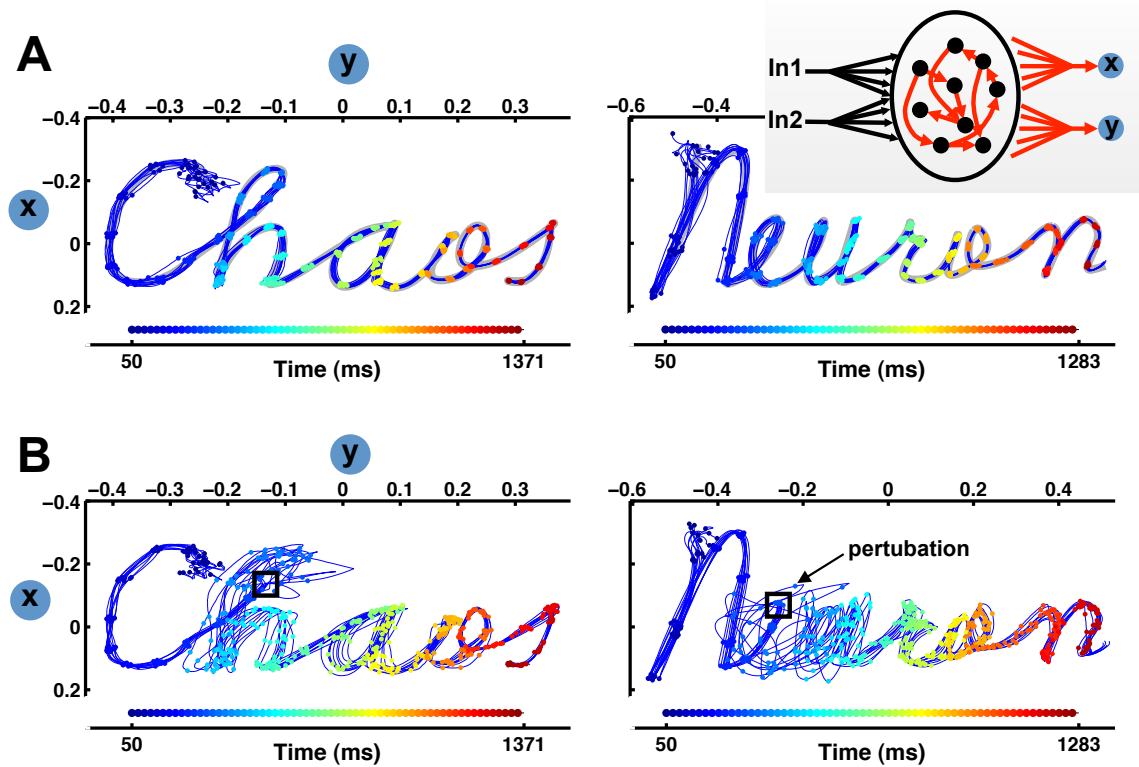

**Figure 5. Generation and stability of complex spatiotemporal motor patterns. A:** Blue traces represent 10 test trials in response to In1 (left panel) or In2 (right) after training; the background gray line shows the output target. These test trials were run over different initial conditions in the presence of continuous noise in all of the 800 recurrent units (max amplitude: $-5 \cdot 10^{-4}$ - $5 \cdot 10^{-4}$). Time is represented by uniformly placed colored circles ($\Delta t \cong 18$ ms). **B:** Test trials run under the same initial condition in the presence of continuous noise, but with the addition of a strong perturbation at 300 ms (open square). The perturbation was produced by an additional 10 ms input pulse (not diagrammed) with an amplitude of 0.2.

To further characterize the computational implications of innate recurrent plasticity to tune RRNs we also simulated a complex spatiotemporal motor task: cursive handwriting. Again two distinct brief inputs (50 ms duration) were used to stimulate an RRN (N=800, g=1.5) in the absence of noise to generate the two innate trajectories for training the RRN. After training the RRN on both trajectories, two output units (representing X and Y axes) were trained and then tested (in the presence of continuous noise) to produce the words "Chaos" and "Neuron" in response to Inputs 1 and 2, respectively (**Fig. 5A**). One of the most remarkable features of creating locally stable trajectories is that they function as a "transient attractors": even relatively large perturbations to the RRN can be self-corrected. This feature is shown by perturbing the network activity after the trajectory has already been initiated. The perturbation was produced by a 10 ms pulse of an additional input randomly connected to all units in the RRN with an input amplitude of 0.2, injected at t=300 ms (corresponding approximately to the time of the *'h'* and *'e'* during the *'Chaos'* and *'Neuron'* trajectories respectively). Despite the obvious effect of the perturbation on the state of the recurrent network (as evidenced by the altered output), the network returned to the original trajectory over the course of a few hundred milliseconds resulting in an increasingly clear writing.

**Figure 6. Innate training dramatically decreases the neural variance in response to stimulus onset.** In order to study the dynamics of trained networks in the presence of very high-levels of noise we used a network with a $p_c$=0.25 (N = 800, g=1.5, 1.3 sec training window), uniformly distributed noise with a range of -0.4 – 0.4 was continuously injected into all recurrent units (noise amplitude is 4.7% of total absolute incoming synaptic weight to a neuron averaged across neurons). As in Fig. 4 the output unit was trained to generate a timed pulse (1000 ms after the onset of the 50 ms input pulse). The upper panel shows the traces of three units over two different trials (blue and red). Note that the effects of the very high noise levels are readily evident in the traces. Furthermore despite the apparent difference in the traces the output unit was able to consistently generate the timed response at approximately 1000 ms (middle panel). The lower panel shows the neural variance. The variance of each unit was calculated over 8 trials, and then averaged over all 800 units. There was a sharp decrease in variance produced by the onset of the stimulus, which persisted over many seconds before gradually ramping back up to baseline (not shown). The dashed line shows the neural variance before training: because the input "clamps" network activity stimulus onset also produced a decrease in the variance, but it rapidly increased after stimulus offset.

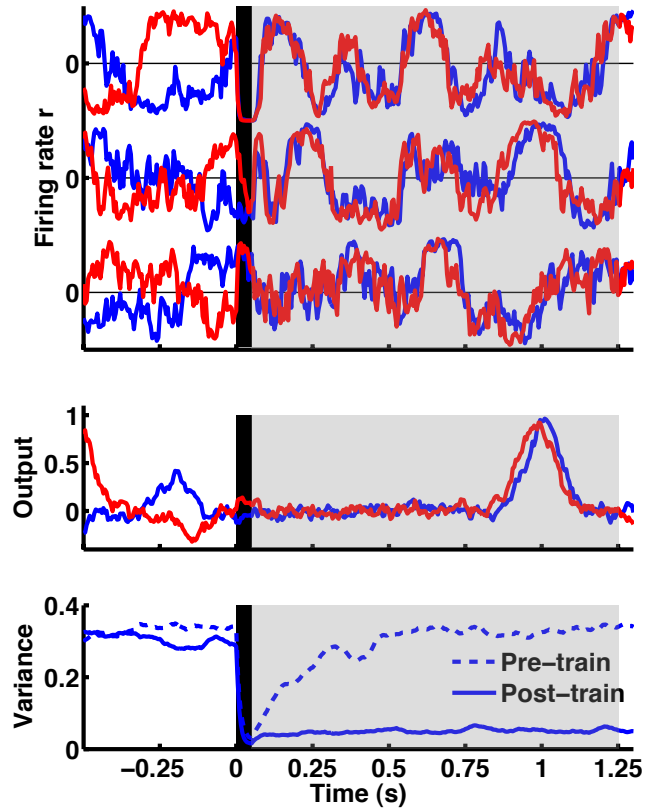

## Experimentally Observed Decreases in Variability

Implicit in the findings above is that after training there are fundamentally different types of dynamics within the same network: while ongoing activity (or trajectories triggered by untrained inputs) continue to produce chaotic trajectories, the trained trajectories exhibit local stable patterns of activity. Recent experimental studies have also revealed different types of dynamics within the same network. For example, it has been shown that cross-trial variability of neural activity is "quenched" in response to stimulus onset (Churchland et al., 2010); that is, the variability of neural "ongoing" or "background" activity is significantly larger than that observed after a stimulus or during a behavioral task. We next quantified the cross-trial variance before and after the brief 50 ms input in the trained and untrained networks. Additionally, to "push the envelope" in terms of how much noise the network can handle we dramatically increased the noise levels. Thus, here the network was trained for 500 trials, as opposed to 20 (30 in Fig. 4). The variance was calculated over 8 test trials for each of the 800 units over a time period starting 500 ms before the stimulus. The target delay was 1000 ms (and the training window was 1300 ms).

The sample firing rates of three units in **Fig 6** (upper panel) show that in presence of continuous very high levels of noise each of the Ex units exhibits significant jitter, reminiscent of the membrane voltage fluctuations observed in vivo, resulting in a high cross-trial variance before stimulation (t<0). In response to the input, a trained network was still able to generate an appropriately timed output (**Fig. 6**, middle panel), despite the readily apparent jitter in firing rates in the individual units. This is a powerful example of how robust a recurrent network can be to very high levels of noise. And as expected, this robustness reflects a dramatic decrease in the variance of the activity after the stimulus onset (**Fig. 6**, lower panel).

**Mechanisms: Network Structure After Training**
As a first approach to understand how training altered the structure of the recurrent networks, we examined the distribution of weights and the connectivity patterns before and after training. The distribution of the nonzero recurrent weights changed very consistently after training (**Fig. 7A**). Innate training led to a non-Gaussian distribution with longer tails (note that the number of nonzero weights does not change because training does not alter which units are connected), meaning that the median absolute synaptic weight became stronger (Pre-train median ± mean absolute deviation from the median, MAD, across 10 networks: 0.1358 ± 0.0004; Post-train: 0.147 ± 0.001; paired Wilcoxon sign-rank test, $p$=0.002). Shuffling the weights (not the connections) of the recurrent matrix $\mathbf{W^{Rec}}$ after training leaves the weight distribution untouched; however, the stability properties of the network are destroyed (**Fig. 7B**). Thus, it's not simply the statistic of the synaptic weights or the binary connectivity what defines the network behavior. As an example of the importance of precise wiring rather than the distribution, we found that post-training weights from bidirectional connections were significantly stronger on average than those from unidirectional connections (in absolute value; unidirectional median ± MAD across networks: 0.145 ± 0.001; bidirectional: 0.161 ± 0.003; paired Wilcoxon sign-rank test, $p$=0.002; see supplemental **Fig. S1**). Interestingly, both the long-tailed weight distribution and the bidirectional vs. unidirectional connectivity features observed here were reported by Song and colleagues (2005) in the rat visual cortex.

In order to explore the role of the connectivity structure of the trained networks we computed the distribution of local clustering coefficients (Watts and Strogatz, 1998). Here we analyzed the presence cyclic clusters, which are associated with recurrency and self-sustained activity, by using a directed, weighted version of the clustering coefficients (Fagiolo, 2007). The cyclic clustering coefficients provide a measure of the number of neuron triplets connected in a circular fashion, weighted by their synaptic strengths. We also analyzed the non-cyclic clusters, in which neuron triplets do not form a closed loop; these motifs could have a role in feedforward propagation of activity (see Methods). As shown in **Fig. 7C**, innate training increased the median cyclic clustering coefficients (Pre-train median ± MAD across networks: 0.01270 ± 0.00005; Post-train: 0.0139 ± 0.0001; paired Wilcoxon sign-rank test $p$=0.002) and made the distribution of the clustering coefficients have a longer right tail (Post-train distributions are non-Gaussian; Lilliefors test, $p$<0.001 for all 10 networks), meaning that the trained networks exhibited stronger short-range recurrency. This result complements a previous report of an increase in recurrency after plasticity within a recurrent network (Liu and Buonomano, 2009). Interestingly, innate training also resulted in an increase in the non-cyclic clustering coefficients (Pre-train median ± MAD across networks: 0.01280 ± 0.00005; Post-train: 0.0142 ± 0.0002; paired Wilcoxon sign-rank test $p$=0.002; see **Fig. 7D**), leading to a stronger short-range feedforward structure.

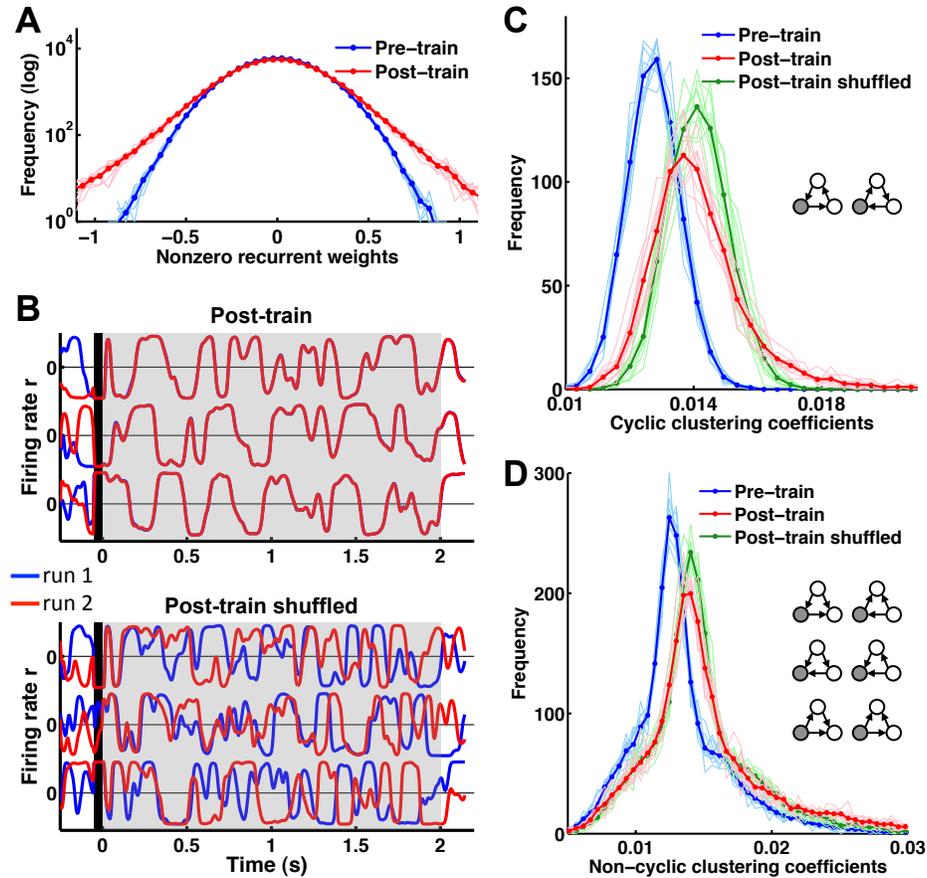

**Figure 7. Effects of training on network structure. A:** Distribution of the nonzero recurrent weights. Thin lines represent the distributions of the weights of ten networks before (blue) and after (red) training. Thick lines represent the averages across the 10 networks. Pre-training: networks are Gaussian by construction. Post-training: all networks are non-Gaussian (Lilliefors test, *p*<0.001 for each of the ten networks). Median absolute synaptic weights significantly increased after training. **B:** Numerical simulation of one trained network before and after shuffling the weights of its recurrent matrix **W^Rec** (two runs each, without noise), showing that the stability properties of the shuffled network are lost despite having the same weight distribution and the same connectivity. **C:** Distribution of local weighted cyclic clustering coefficients. Training leads to an increase in the cyclic clustering coefficients. Shuffling (green) of the weights of the Post-train recurrent matrix **W^Rec** significantly changed the cyclic clustering distribution. **D:** Distribution of local weighted non-cyclic clustering coefficients. Training also increased the median non-cyclic clustering coefficients.

To determine whether the observed dynamics reflected the specific wiring signature of the trained networks, we calculated both cyclic and non-cyclic clustering distributions after shuffling the weights (not the connections) of the trained networks (**Fig. 7C** and **D**). Shuffling significantly altered the distribution of the cyclic distribution more than that of the non-cyclic coefficients (two-sample Kolmogorov-Smirnov test between Post-train and Post-train Shuffled for every network, all p values < 0.002; for the non-cyclic distributions: *p* values ranged from 0.002 to 0.11), suggesting that the presence of cyclic clusters may have an important role in the ability of an RRN to generate complex yet stable neural trajectories. However, as we showed above an untrained input can produce a chaotic trajectory after training, and thus it should be stressed that some interaction between the input and the structure of the recurrent network is involved in the resulting dynamics.

## DISCUSSION

A fundamental finding described here is that with the appropriate tuning of recurrent synapses, initially chaotic trajectories can be transformed into locally stable ones. These locally stable regimes have profound computational implications and reveal that neural networks can exhibit two different self-generated modes of neural dynamics. Previous theoretical work established that recurrent networks in high-gain regimes (that have self-sustaining activity) can generate complex, but unstable and chaotic trajectories (Brunel, 2000; Sompolinsky et al., 1988; van Vreeswijk and Sompolinsky, 1996). We now demonstrate that through recurrent synaptic plasticity stable trajectories can be "burnt in", these "learned" trajectories are locally stable over many seconds despite the fact that all units in the network have a 10 ms time constant. These second long stable trajectories are, of course, in the range of most behavioral tasks, and potentially account for a large body of neurophysiological data showing reproducible patterns of activity across trials despite high levels of noise and evidence of chaos. The computational power of these recurrent networks with stable trajectories is demonstrated by their ability to drive the activity of output units capable of generating timed responses or complex motor patterns (handwriting).

### Plasticity Within Recurrent Networks

We provide one of the first examples showing that synaptic plasticity within recurrent networks in a high-gain regime can dramatically enhance their computational power. While many models have, of course, incorporated learning rules into recurrent network, few of these models operate in high-gain regimes capable of self-perpetuating activity. Some studies have incorporated more realistic learning rules in recurrent networks, but these are only transiently active or are not robust to high noise (Fiete et al., 2010; Izhikevich, 2006; Liu and Buonomano, 2009). And when plasticity has been successfully incorporated into models similar to that examined here, it did not enhance performance. For example, in the work that inspired the current approach, Sussillo and Abbott (2009) incorporated plasticity within the recurrent network, which was guided by the error in the output units, but they did not observe any significant computational advantage of recurrent plasticity compared to plasticity of the feedback units alone (see also, Sussillo and Abbott, 2012).

While the current study demonstrates the power of using plasticity to tune recurrent weights, it is important to emphasize that the specific learning rule and strategy outlined here is highly supervised, and is not biologically realistic. First, although the RLS rule is "delta rule-like" in that it minimizes an error, it is computationally sophisticated and as applied here operates on a unrealistically fast time scale—however, as previously noted there may be conditions under which more plausible rules can be used (Sussillo and Abbott, 2009). Second, in our implementation there is a separate target pattern that guides plasticity for each unit in the network: a highly implausible biological scenario. Nevertheless, in one sense the rule is more biologically plausible than traditional supervised learning rules: the recurrent plasticity does not actually require any externally imposed "desired" target to guide plasticity because the network learns its "innate" target. That is, there is no external teacher that dictated the "correct" target pattern, because the innate trajectory was obtained internally. Thus, more realistic learning rules may be viable because which trajectories are "burned-in" is largely irrelevant, what matters is that some arbitrary subset of innate trajectories is learned. Thus future research will have to address whether similar regimes can be achieved with biologically plausible learning rules.

The ability to modify the recurrent weights RRNs has immediate implications for computational neuroscience and reservoir computing approaches as it allows networks to harness the computational potential of the complex dynamics of RRNs while avoiding the sensitivity to noise such systems generally exhibit (Jaeger and Haas, 2004; Jaeger et al., 2007; Sussillo and Abbott, 2009). A large number of models have proposed that the complex activity patterns generated by recurrent networks maybe used to represent complex sensory stimuli or generate motor patterns (Buonomano and Laje, 2010; Buonomano and Mauk, 1994; Buonomano and Merzenich, 1995; Haeusler and Maass, 2007; Jaeger, 2001; Jaeger and Haas, 2004; Karmarkar and Buonomano, 2007; Maass et al., 2002; Mauk and Donegan, 1997; Medina et al., 2000). Within this framework the RRN functions as a "reservoir"; for example, any complex and reproducible pattern of activity can be used to tell time by teaching an output unit to recognize the pattern of activity in the RRN that corresponds to the target time point (Mauk and Donegan, 1997; Medina et al., 2000). The limitation of these approaches to date has been that the RRNs (the "reservoir") are generic, they do not adapt to the task at hand. It has been difficult to overcome this hurdle because of the inherent challenges of changing the weights that are used over and over again within the same trajectory—e.g., modifying a synaptic weight because the network is not in the desired state at t=200 may change the entire trajectory of the network starting at t=0 ms at the next trial.

**Structure and Mechanisms of Underlying Stable Trajectories**

In linear networks the structure of recurrent networks, as analyzed though a number of techniques, including eigen and Schur decompositions, provide valuable methods to understand and describe the dynamics of such systems (e.g., Goldman, 2009). However, predicting the behavior of a continuous-time, high-dimensional nonlinear network from its connectivity matrix is still not possible in the general case. Additionally, a key observation here is that the interaction between the input connectivity and recurrent weights plays a fundamental role in how the network responds to external stimuli: as shown here after training the same network can respond very differently to different inputs (Figs. 2, 3, 5). Steps towards understanding this interaction and the dynamics in response to external inputs have been taken for both discrete-time linear networks (Ganguli et al., 2008) and continuous-time nonlinear networks (Rajan et al., 2010), but the fact remains that in continuous nonlinear networks it remains impossible to predict the modes of activity or describe why some trajectories are locally stable and others are not.

Despite the limitations in mathematically analyzing and predicting the dynamics of nonlinear networks, it is of interest that analysis of the connectivity patterns and network structure revealed highly reproducible, non-random signatures in the recurrent weight matrices. For example, innate training produced a robust increase in the median absolute weight resulting in a non-Gaussian long-tailed weight distribution (**Fig. 7A**). Additionally, training produced a larger increase in the strength of reciprocal connections than unidirectional connections. Interestingly both these features have been observed in the connectivity between neocortical pyramidal neurons (Cheetham et al., 2007; Song et al., 2005). Finally, another very robust effect of training was a change in the distributions of the cyclic clustering coefficients. In particular the distribution of cyclic cluster coefficients changed to a highly non-random one, suggesting that a different short-range recurrent circuitry than that observed in RRN with randomly assigned weights might contribute to the generation of stable trajectories.

## Computational Implications and Experimental Predictions

One key prediction that arises from the current paper is that recurrent cortical circuits exhibit "preferred" or learned neural trajectories. That is, while spontaneously active networks exhibit complex but unstable trajectories, familiar stimuli elicit "preferred" stable trajectories that can last many seconds and are highly robust to noise. The presence of two modes of activity is consistent with experimental evidence. Specifically, while networks can exhibit ongoing and highly variable spontaneous activity, in response to a stimulus specific trajectories are elicited, and these have a much lower cross-trial variability than spontaneous background activity (Churchland et al., 2010). We show (**Fig. 6**) that the networks studied here reproduce the experimentally observed decrease in neural variance in response to stimulus onset. A stronger and testable experimental prediction, however, is that the magnitude of the variance drop and its duration is stimulus specific and dependent on learning. That is, the decrease in variance in response to overtrained stimuli will be larger and longer-lasting than that to novel or irrelevant stimuli.

Our results demonstrate that, in principle, recurrent plasticity can locally suppress chaos and dramatically enhance the computational power of recurrent networks. Specifically, we provide an example of how trajectories of recurrent networks can be used to time events in the presence of significant levels of noise, and generate complex time-varying motor patterns (**Fig. 4, 5**). A novel feature of our approach is the notion of "dynamic" or "transient" attractors, which account for the ability of a network to not only generate the hand-written patterns of **Fig. 5**, but to be able to return to the pattern in response to large perturbations. To the best of our knowledge this is the first description of a high-dimensional nonlinear system capable of this level of robustness.

The current findings may also shed light on a longstanding puzzle in neuroscience: theoretical and experimental studies suggest that the brain exhibits chaotic regimes (Banerjee et al., 2008; Brunel, 2000; Izhikevich and Edelman, 2008; London et al., 2010; Skarda and Freeman, 1987; van Vreeswijk and Sompolinsky, 1996); yet experimental evidence, and common sense, also tell us that neural circuits can generate reproducible neural trajectories critical for sensory and motor processing (Churchland et al., 2012). Our description of a system that has coexisting complex stable transients and chaotic trajectories may reconcile the existence of these two regimes within the brain.

## ACKNOWLEDGMENTS

We thank Alan Garfinkel and Ramón Huerta for helpful discussions and comments on the manuscript. This work was supported by the National Institute of Mental Health (MH60163), the National Science Foundation (II-1114833), the Pew Charitable Trusts, and CONICET (Argentina).

**SUPPLEMENTAL TEXT**

**Comparison to other chaos-related regimes**

Below we briefly summarize other chaos-related regimes in complex systems to provide a contrast between different related phenomena. To the best of our knowledge a critical difference between our results and all previous known regimes is that the stable transient channel demonstrated here is the result of explicit modifications of the system itself— modifications designed to create stable trajectories within a chaotic system. This contrasts with previous work, which describe "natural" regimes of intact systems.

*Regular chaos*. Although many of our RRNs retained a chaotic attractor after training, the fact that the phase space has some locally stable trajectories makes our RRNs different from a regular chaotic system. In a regular chaotic system, the divergence rate of trajectories naturally fluctuates along the trajectory, and thus finding a particular portion of a trajectory where the flow converges rather than diverges is possible (Kantz, 1994); however, the probability of finding a long stable trajectory is very low. The convergence displayed at very short times (<50 ms) in **Fig. 2A** is only due to the fast relaxation of a random initial condition towards the attractor (Wolf et al., 1985), and can also be found in a regular chaotic system.

*Stable Chaos*. Among the different type of regimes observed in complex and chaotic systems *"stable chaos"* deserves special mention because it has been studied in randomly connected recurrent networks composed of integrate-and-fire units (Politi et al., 1993; Zillmer et al., 2009). In this context *"stable chaos"* has been used to refer to irregular transient behavior despite a negative LLE. In these regimes periodic solutions are generally reached after a very long transient that is very irregular. Although the largest non-zero Lyapunov exponent is negative (i.e., the system is stable under infinitesimal perturbations), finite-size perturbations can lead to diverging trajectories, what is reflected in intricate basins of attraction for the different periodic solutions. There are a couple of differences between this phenomenon and our finding. First, we observe locally stable trajectories in response to large perturbations (e.g., Figure 2, 5, 6). Second, in most cases the system described here does not converge to a limit cycle and thus remains formally chaotic outside the training window or in response to different inputs.

For a second and distinct use of the term *stable chaos* see Milani and colleagues (Milani and Nobili, 1992; Milani et al., 1997).

*Strange nonchaotic attractors*. First described by Grebogi et al (1984), certain types of dynamical systems display solutions that are complex but do not show sensitive dependence on initial conditions. In these systems, the attractor is a geometrically complex object (e.g., it has a fractal structure), which leads to complex trajectories in phase space and thus to complex time series. Two initially close trajectories, however, don't diverge exponentially in time and thus the LLE is non-positive. Although some features of our trained trajectories might resemble those of strange nonchaotic attractors (namely, complex time series and a non-positive LLE), there are critical differences: our stable trajectories are transient, whereas a strange nonchaotic attractor is an invariant solution (i.e., once the system is in the attractor it will remain in the attractor) with well-defined stationary statistical properties. A second, important difference is that in our case the LLE after training is non-positive for the trained trajectory only—for many networks, the rest of the phase space retains a positive LLE.

***Chaotic transients***. Under certain circumstances, the transient evolution of a system before reaching a non-chaotic stable solution can be chaotic itself, what is called "transient chaos" (Grebogi et al., 1983). This phenomenon is in many way the opposite of the stable transients we report: in transient chaos the LLE is locally positive and in our networks the LLE is locally non-positive.

## SUPPLEMENTAL REFERENCES